\documentclass[12pt,preprint]{aastex}
\usepackage{amsmath}
\usepackage{xcolor}
\usepackage{bm}
\usepackage{url}
\usepackage{lineno}
\definecolor{rkka}{RGB}{219,66,32}

\shorttitle{Astrometric mass ratios of 248 binary stars}

\newcommand{\gbp}{$G_{\rm BP}$}
\newcommand{\grp}{$G_{\rm RP}$}
\newcommand{\eb}{\begin{equation}}
\newcommand{\ee}{\end{equation}}
\newcommand{\norm}[1]{\left\lVert#1\right\rVert}

\begin{document}

\title{Astrometric mass ratios of 248 long-period binary stars resolved in Hipparcos and Gaia EDR3}

\author{Valeri V. Makarov}
\affil{U.S. Naval Observatory, 3450 Massachusetts Ave., Washington, DC 20392-5420, USA}
\email{valeri.makarov@gmail.com}
\author{Claus Fabricius}
\affil{Departament de FQA, Institut de Ci\`encies del Cosmos, Universitat de Barcelona (IEEC-UB), 
Mart\'i i Franqu\`es 1, 08028 Barcelona, Spain}
\email{claus@fqa.ub.edu}

\begin{abstract}
Using the absolute astrometric positions and proper motions for common stars in the Hipparcos and Gaia catalogs
separated by 24.75 years in the mean epoch, we compute mass ratios for long-period, resolved binary systems
without any astrophysical assumptions or dependencies  except the presence of inner binary subsystems that may perturb the observed mean proper motions. The mean epoch positions of binary companions from
the Hipparcos Double and Multiple System Annex (DMSA) are used as the first epoch. The mean positions and proper motions of carefully cross-matched counterparts in Gaia EDR3 comprise the second
epoch data. Selecting only results with sufficiently high signal-to-noise ratio and discarding numerous optical pairs, we construct a catalog of 248 binary systems, which is published online. Several cases with unusual 
properties or results are also discussed. 
\end{abstract}

\keywords{
astrometry --- binaries: visual --- stars: fundamental parameters}

\section{Introduction}
\label{int.sec}

About 20\% of stars in the ground-breaking Hipparcos catalog \citep{esa} are components of double or multiple systems. Those that are sufficiently bright and well-separated were treated with ad hoc data reduction
procedures designed to estimate a larger number of unknowns than the standard 5-parameter model. For example,
a resolved double without a measurable orbital motion was typically treated in a 7-, 9-, or 10-parameter
adjustment solution, depending on the assumptions about the proper motion and parallax of the secondary
component. The resulting astrometric fit is present in the main catalog for the primary (nominally, brighter)
components, while the secondaries are, in most cases, missing there. The complete catalog of these special
solutions is available in the Double and Multiple System Annex (DMSA). This information about double stars
can also be found in a more user-friendly and streamlined form in the Tycho Double Star Catalog \citep[TDSC,][]{2002A&A...384..180F}, complemented with numerous additional systems resolved in the Tycho mission.
The Gaia mission, on the other hand, has not yet released specialized solutions for close double or multiple
systems \citep{pru,bro}. Such objects have been treated in the same way as single stars in the general data reduction
pipeline. Many double stars have been resolved in the current Gaia EDR3, however, owing to the high
dynamical range and good angular resolution of the instrument. Significant astrometric perturbations of
binary stars processed with the general 5- or 6-parameter models have already been reported for Gaia DR1
\citep{mafa}.

Wide binaries with periods longer than 100 yr require decades long ground-based observations to derive
their apparent orbits. Most of this work relies on the methods of differential astrometry where only the relative
position of the secondary with respect to the primary is involved. Combined with a trigonometric parallax,
the estimated orbit provides the total mass of the binary system. The distribution of mass between the components
is much harder to estimate. Most of these estimates are based on the photometric data (relative brightness and color) combined with models of stellar evolution. The most accurate empirical masses (and mass ratios)
are obtained for the rather rare class of spectroscopic double-lined binaries that are close and bright
enough for precision astrometry \citep{2002AJ....124.1716T}. 
In this paper, we consistently apply the method of absolute
two-epoch astrometry \citep{mak21} to the entire collection of doubles in the DMSA cross-matched with Gaia EDR3.
The goal is to select likely physical binaries with a measurable orbital motion between the mean epochs (1991.25
for DMSA and 2016 for EDR3) and to estimate their mass ratios along with their formal uncertainties. 
The main requirements for this method to work are that the orbital period should be
significantly longer than the duration of each astrometric mission, which is about $3.5$ years for Hipparcos
and 33 months for Gaia EDR3, and the system
should be close enough to produce measurable changes in the apparent orbital motion of the components.

Much of the effort on this route was spent on constructing the initial sample of suitable double and triple
systems, which is described in Section \ref{sel.sec}. The resulting collection of 6306 cross-matched and 
verified Hipparcos$/$EDR3 double systems was processed with a mass-ratio estimation and Monte Carlo
simulation algorithms (Section \ref{q.sec}). Additional vetting based on objective signal-to-noise and
model consistency criteria reduced the sample to 248 binaries, for which detailed information from Hipparcos,
Gaia, and our own results is collected in a catalog published online (Section \ref{res.sec}). Objects of
special interest with outstanding or unexpected properties are selectively discussed in Section \ref{obj.sec}.
Discussion of future development and possible applications is presented in the concluding Section \ref{con.sec}.

%%--------------------------------------------------------------
\section{Selection of resolved binary systems}
\label{sel.sec}
To generate a sample of cross-matched double and triple systems in Hipparcos and Gaia EDR3, we begin with the
Double and Multiple System Annex to the main Hipparcos catalog \citep{esa}, which lists 12005 double, 182 triple, and 8 quadruple star systems. The components of these systems are not necessarily physically bound---in fact, many of them prove to be optical combinations, i.e., chance projections of unrelated stars on the sky. These
sources were treated by a special pipeline separately from the regular 5-parameter solutions. The data reduction
algorithm is much more complicated and the results are less reliable because of the greater number of unknowns
and the entangled character of the primary observations. Due to the periodical character of the recorded photon
counts, multiple solutions to the same set of modulated curves can be found, and selecting the correct solution
requires accurate prior knowledge about the absolute and relative position of the components \citep{2000A&AS..144...45F}. Having one of the components moving by measurable distances with respect to the other (as in optical pairs due to the difference in parallax or proper motion or in nearby binaries due to orbital motion) or significantly varying in brightness exacerbates the problem to the point of being unsolvable. For this reason,
most of the solutions in DMSA are flagged as ``fixed", i.e., they were constrained to have the same proper
motion and parallax. The effect of fixed solution for optical pairs is especially damaging, because the mean positions are determined at the standard epoch 1991.25, and the unaccounted proper motion differences may compound
for fast-moving components whenever the true optimal epoch of observations minimizing the trace of the covariance
matrix deviates from this common epoch.

Treating each DMSA component as a separate entity, we found 14793 counterparts in Gaia EDR3 by a cone search\footnote{For this initial cross match we used the match facility included in the Gaia archive at \url{https://gea.esac.esa.int/archive/}} with a radius of $2.5$ arcsec around positions propagated to the Gaia mean epoch with Hipparcos proper motions. In cases of multiple counterparts within the search cone, the closest entry was selected as the likeliest match. Subsequently, each recovered Gaia entry was matched to the nearest DMSA component. This procedure left
234 DMSA entries without any Gaia counterparts. To identify the reason for these misses, we carefully
investigated a smaller selection of 12 pairs from this lot. Only a small fraction (less than 10\%)
are truly missing in Gaia because they are too bright and saturate the detector pixels even in the
special gating mode. The majority of companions are present in Gaia
EDR3 but outside of the search radius. Many of them are parts of optical pairs (chance projections) with
significantly different proper motions and, possibly, parallax, which obtained ``fixed" solutions with a common proper motion in the DMSA processing. The examined 12 cases are predominantly widely separated pairs up
to $26\arcsec$ with much degraded Hipparcos proper motions. Gross errors in position are also present.
In 3 cases out of 12, greatly improved astrometric solutions were obtained with more accurate initial assumptions
by \citet{2000A&AS..144...45F} using the published Hipparcos transit data. A larger number of difficult
cases can be revisited today utilizing additional constraints from Gaia.

The distribution  of distances between the matched counterparts is strongly peaked at about $0.03$ arcsec, which
corresponds to the expected accumulated proper motion error. A long tail of this distribution stretches
all the way to $2.5$ arcsec. A total of 34 components (all B or C in DMSA) are more distant than 2 arcsec
from their matched  Gaia counterparts. Almost all of them appear to be optical pairs with a few gross errors
in DMSA, such as HIP 75741 where the true companion is as faint as 12.3 mag and located on the other side
of the primary. Another few systems appear to be genuine binaries where the secondary components moved long
distances on the sky in the intervening 24.75 years, such as HIP 13716, 43225, and 117163.

Further pruning of the sample was aimed at removing Gaia entries with a flag indicating that no parallax
or proper motion was available (1125 partial solutions). This elimination broke up additional systems. We
identified 773 DMSA primaries that had lost their secondaries and removed them. An additional 75 surviving
secondaries without any primaries were then found and removed. Among the remaining 74 tertiary components,
1 was found to be lacking a secondary, and that system was dropped. The remaining 12818 components include
4 original DMSA quadruple systems represented by 15 cross-matched components, namely, HIP 4121 ABCD, 23624
ABCD, 26020 AB $+$ 26026 C, and 26221 A $+$ 26220 BD $+$ 26224 C. These were carefully examined, determined
to be optical systems, and rejected.

At this point, 191 components of original DMSA triple systems remained in the sample. This paper concerns only binary stars, so we removed all the triple systems. The emerging working sample includes 12612 components of 6306
cross-matched binaries. This data set is technically suitable to apply our mass-ratio  calculation algorithm,
and a heavy filtering and cleaning of the results is still in order.

\section{Calculation of astrometric mass ratios}
\label{q.sec}
Details on the mass ratio determination from absolute astrometry of two epochs can be found in
\citet{mak21}. Here we only provide a brief summary of the algorithm for the sake of presentation
consistency. A resolved binary star with two components $i=1,2$ have position unit-vectors $r_{ij}$ in two
astrometric catalogs, $j=1,2$, at epochs $t_1$ and $t_2$. We compute the long-term proper motion
$\nu_i=(r_{i2}-r_{i1})/\Delta t$ for each component, where $\Delta t$ is the epoch difference $t_2-t_1$.
The second epoch catalog (Gaia EDR3) provides short-term proper motions $\mu_{i2}$. Following directly from the
inertial motion of the center of mass in space, without any implicit astrophysical assumptions or
data, the mass ratio $q=m_2/m_1$ may be computed as
\eb
q=\frac{\norm{\mu_{12}-\nu_1}}{\norm{\nu_2-\mu_{22}}}.
\label{nor.eq}
\ee
The vectors $v_1=\mu_{12}-\nu_1$ and $v_2=\nu_2-\mu_{22}$ should ideally be aligned in the tangential plane. Random and systematic measurement errors, as well as physical perturbations, which we will discuss later, make these
observed vectors nonparallel. The degree of misalignment, defined as
\eb\eta= \arccos{\left(v_1\cdot v_2/(\norm{v_1} \norm{v_2})\right)},
\label{eta.eq}
\ee
is an important
indicator of the reliability and quality of a specific estimation.

These formulae were used for the sample of 6306 cross-matched and verified Hipparcos$/$EDR3 double systems
described in the previous Section. Computations of $q$ and $\eta$ are straightforward and fast, but the
resulting values should be accompanied by formal errors to evaluate their uncertainties. To this end,
we perform 4000 Monte Carlo simulations perturbing the observed positions and proper motions for each component
in accordance with the full covariance matrices of the involved astrometric parameters, as detailed in \citet{mak21}. The 0.1573 and 0.8427 quantiles of the emerging sampled distributions of $q$ are the estimated
boundaries of the 0.6854 confidence interval, which corresponds to the $\pm 1\sigma$ confidence interval for
a Gaussian distribution of probabilities. The differences between these quantiles and the median $q$ are called the lower bound error
($e_{q-}$) and upper bound error ($e_{q+}$) in this paper.

Apart from the angle $\eta$ and the interval of uncertainties, we found a quantity called SNR (signal-to-noise
ratio) quite useful for separating the wheat from the chaff. Even though most of the working sample are genuine
widely separated binaries of considerable astronomical interest, they are too distant from us 
and moving too slowly on the sky to generate a measurable proper motion signal over $\Delta t=24.74$ years.
This value is computed for each pair as
\eb
{\rm SNR}={\rm Min}\left[ v'_i\cdot C_i^{-1} \cdot v_i\right]^\frac{1}{2}, \qquad i=1,2,
\label{snr.eq}
\ee
where $C_i$ is the full $2\times2$ covariance matrix of component $i$. This is a multivariate analog of the
commonly used estimate of a derived quantity as a multiple of its estimated standard deviation.

\begin{figure}
\plotone{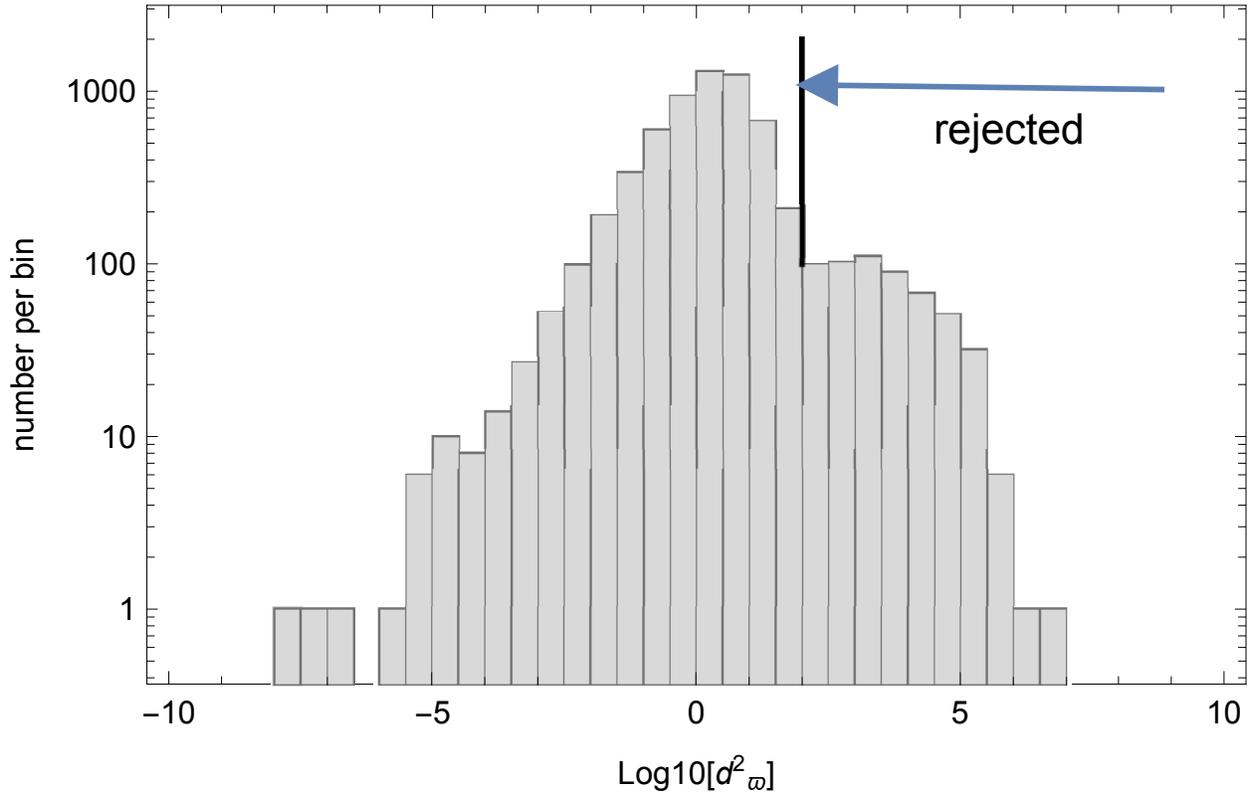}
\caption{Histogram of the decimal logarithm of normalized parallax differences for 6306 resolved and cross-matched doubles in Gaia EDR3 and Hipparcos. The vertical bar shows the initial selection threshold to remove
likely optical pairs.
\label{logpar.fig} }
\end{figure}

To identify likely optical pairs, we compute a quantity called normalized parallax difference, which is
\eb
d_\varpi=\left[(\varpi_2-\varpi_1)^2/(\sigma^2_{\varpi 1}+\sigma^2_{\varpi 2})\right]^\frac{1}{2},
\ee
using EDR3 parallaxes and their formal errors ($\sigma$). The histogram of the decimal logarithm of $d_\varpi^2$
for the entire working sample is shown in Fig. \ref{logpar.fig}. There is a dip separating the main
peak that contains mostly genuine physically bound components from the secondary bump representing optical
pairs. The dip is shallow, so no matter where we select the threshold $d_\varpi$, a small fraction of
optical pairs will contaminate the sample. We choose a rather relaxed limiting value $d_\varpi=10$ taking into
account that the quality of Gaia parallaxes may be degraded for double stars. The adverse effects of duplicity
on Gaia astrometry starting from the first release are evident \citep{mafa, are, fab} but cannot be corrected or
calibrated, and the best strategy is to give accommodation for the increased dispersion if possible.

A more radical cut to the starting sample is provided by the SNR estimates (Eq. \ref{snr.eq}). The orbital motion
of distant binaries is too slow to generate a measurable difference between the long-term and short-term
proper motions. Preliminary results reveal a strong pile-up of estimated $q$ at zero, which is caused by such
distant systems dominating the sample. The likely reason for this bias is the stronger perturbation of
astrometric solutions for the secondaries in Gaia. This perturbation is also confirmed by a large fraction of
estimates with misalignment angles $\eta$ dispersed across the range $[0,\pi]$. For the initial cleaning
step, we choose a limit of SNR$>\sqrt{5}$. Together with the $d_\varpi<10$ filter, this eliminated most of the sample
with only 884 binaries remaining.

This sample still includes a substantial population of binaries with unreliable results, as witnessed by a
widely dispersed angles $\eta$. Both Hipparcos and especially Gaia EDR3 have limited resolution power
on double stars. The observational sample is dominated by wide binaries with a small amount of orbital motion
corresponding to our timeline of 24.75 years. As a filter for our next reduction, we use a rough proxy
of the orbital period:
\eb
P_{\rm proxy}=\left(\frac{\rho}{|\varpi|}\right)^\frac{3}{2},
\ee
with $\rho$ being the angular separation of the components in Gaia EDR3 in mas, $\varpi$ the parallax from
Gaia EDR3 in mas (the absolute value is required because there are negative parallaxes at this point).
$P_{\rm proxy}$ is equal to the orbital period for a total mass of 1$M_{\sun}$ of the binary when the
the projected separation equals the semimajor axis. The actual orbital period can be longer because $\rho$ is smaller than the instantaneous separation in 3D, or it can be shorter
because the total mass is likely to be greater than $1\,M_{\sun}$. In the statistical sense,
the median ratio $\rho/a$ is close to unity for a thermal distribution of eccentricity \citep{2020AJ....160..268T},
but the sample distribution of estimated periods is skewed toward longer values.
The calculated distribution of $\log_{10}
[P_{\rm proxy}]$ appears to include two overlapping bumps, the sharper one being centered at 2.7 (several hundred years) and the wider one at $\sim4$. Our epoch difference is too short to be hoping to catch binary
systems with long orbital periods in motion. We apply an additional filter on $\log_{10}[P_{\rm proxy}]<3.4$ leaving 422 binaries.

The final round of trimming directly uses the calculated $\eta$, as well as a stricter limit on the SNR. A combination of $\eta<50\degr$ and SNR$>3.3$ yields the final sample of 248 binary systems with estimated
mass ratios, which is published with this paper.

\section{Description of the results}
\label{res.sec}
The table of results (published online only) includes 248 rows. Each row corresponds to one unique binary and
includes information from Hipparcos, Gaia EDR3, and our computed parameters for both components. The first four columns contain the Hipparcos identification numbers and component names (capital letters) for the primary and secondary components. The identification numbers are not always identical for binary system components in
Hipparcos. The letters can be A, B, C, D, and S. Columns 5 through 20 contain Gaia EDR3 information about
the primary component, including its source id, RA and Dec coordinates, parallax, proper motion, formal errors of
the five astrometric parameters, astrometric $\chi^2$, RUWE, and $G$, \gbp, \grp\ magnitudes (when available).
The third block of columns (21 -- 36) includes the same information about the secondary component. The remaining columns provide data derived in this paper: $q=m_2/m_1$, alignment angle $\eta$ in degrees, lower bound error
$e_{q-}$, upper bound error
$e_{q+}$, median $q$ from Monte Carlo simulations, SNR, separation in Gaia in mas, position angle from Gaia in degrees, quality flag $\{1,2,3\}$. The highest quality solutions (flag 1) are those with SNR$>50$ and
$\eta<7.45\degr$, the lowest quality (flag 3) is assigned to entries with SNR$<50$ and
$\eta>7.45\degr$, and the quality 2 comprises the remaining solutions. The median errors of $q$
on both sides, $(e_{q-},e_{q+})$, are for quality 1: $(-0.052,+0.057)$, for quality 2: $(-0.124,+0.135)$,
for quality 3: $(-0.183,+0.217)$. Quality 1 through 3 subsets comprise 80, 91, and 77 binaries, respectively.
The median parallaxes are 27.9 mas for quality 1 systems, 18.8 mas for quality 2, and 14.6 mas for quality 3.

\begin{figure}
\plotone{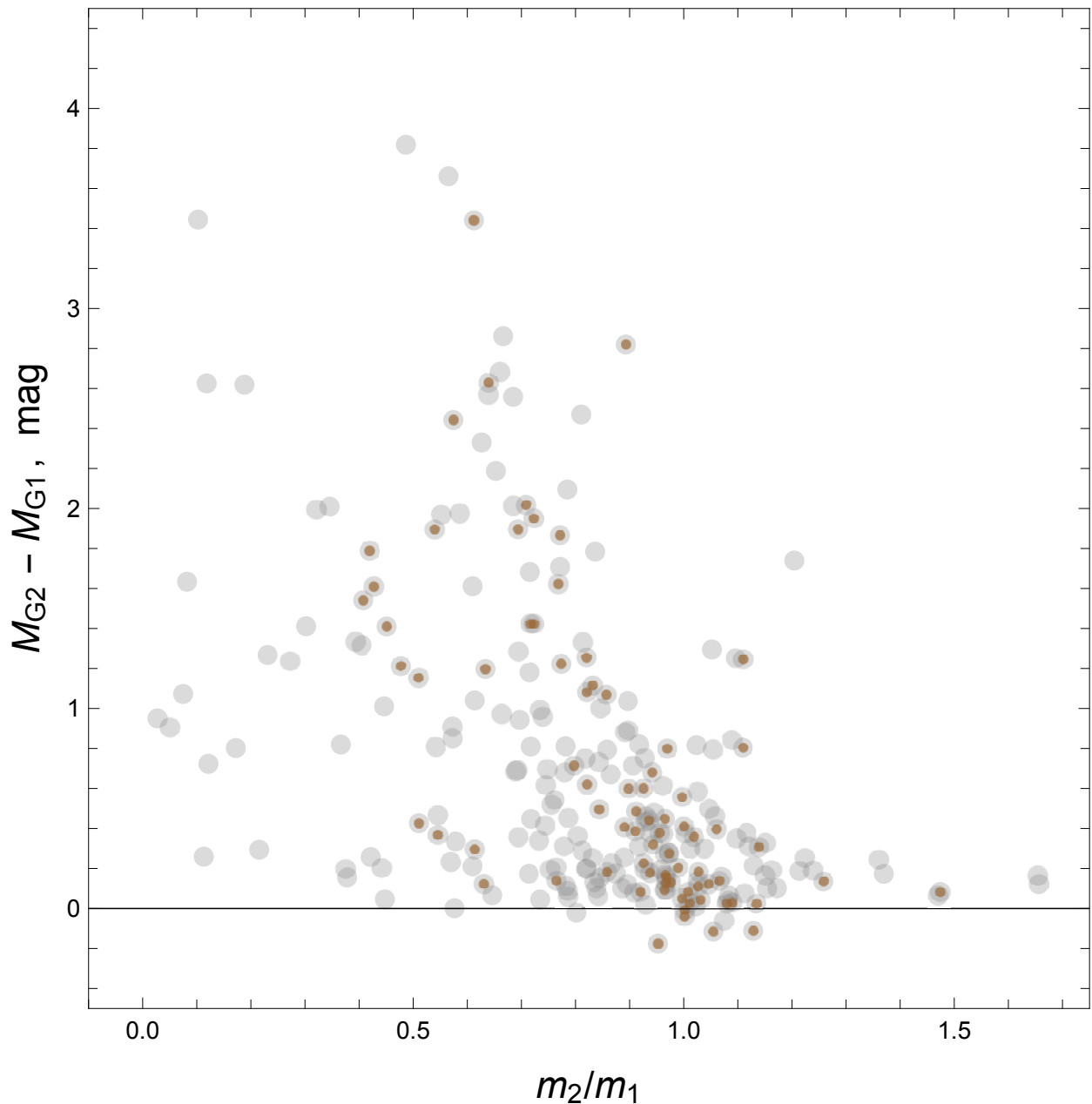}
\caption{Absolute $G$ magnitude difference between the components of 248 binary systems versus their estimated
mass ratio. The entire sample is represented with semitransparent gray circles, while smaller brown dots
show only the quality 1 solutions.
\label{deltamag.fig} }
\end{figure}

As a sanity check, we compute the absolute $G$ magnitudes for all 496 components using EDR3 parallaxes
and observed $G$ magnitudes. The result is depicted in Fig.~\ref{deltamag.fig}. The great majority of the
sample are found within a steep upper envelope as expected. The envelope extends to $q>1$, however, where
even some quality 1 pairs reside. We will see that not all of these cases of fainter and apparently more
massive secondaries are caused by errors in the data. Many, if not all, cases of high-quality solutions
in this category are caused by unresolved triple and higher-multiplicity systems. The effect of
close companions is twofold. They can significantly perturb the short-term proper motions in both astrometric
catalogs. 
Hidden companions to the resolved secondaries also boost their dynamical masses to a much greater degree than
their brightness. These options are impossible to disentangle with the data at hand. We conclude that pairs with
$q>1$ and fainter secondaries are prime candidates for hierarchical multiplicity. A few pairs with $q>1$ and
slightly brighter secondaries are very likely swapped designations in the DMSA for near twins, which are
rather inevitable when one or both of the components are variable. Theoretically, near-twin
components can change their relative brightness in $G$ and $Hp$ magnitudes because of the differences in
the spectral transmissivity curves and some peculiar colors. Using the photometric polynomial transformation for
$G-Hp$ as a function of \gbp$-$\grp\ color\footnote{\url{https://gea.esac.esa.int/archive/documentation/GEDR3/Data_processing/chap_cu5pho/cu5pho_sec_photSystem/cu5pho_ssec_photRelations.html}}, we made sure that there is only one pair in our catalog with the expected magnitude differences $G_1-G_2$ and $Hp_1-Hp_2$ of opposite signs,
namely, the near-twin system HIP 99689 AB.

\begin{figure}
\plottwo{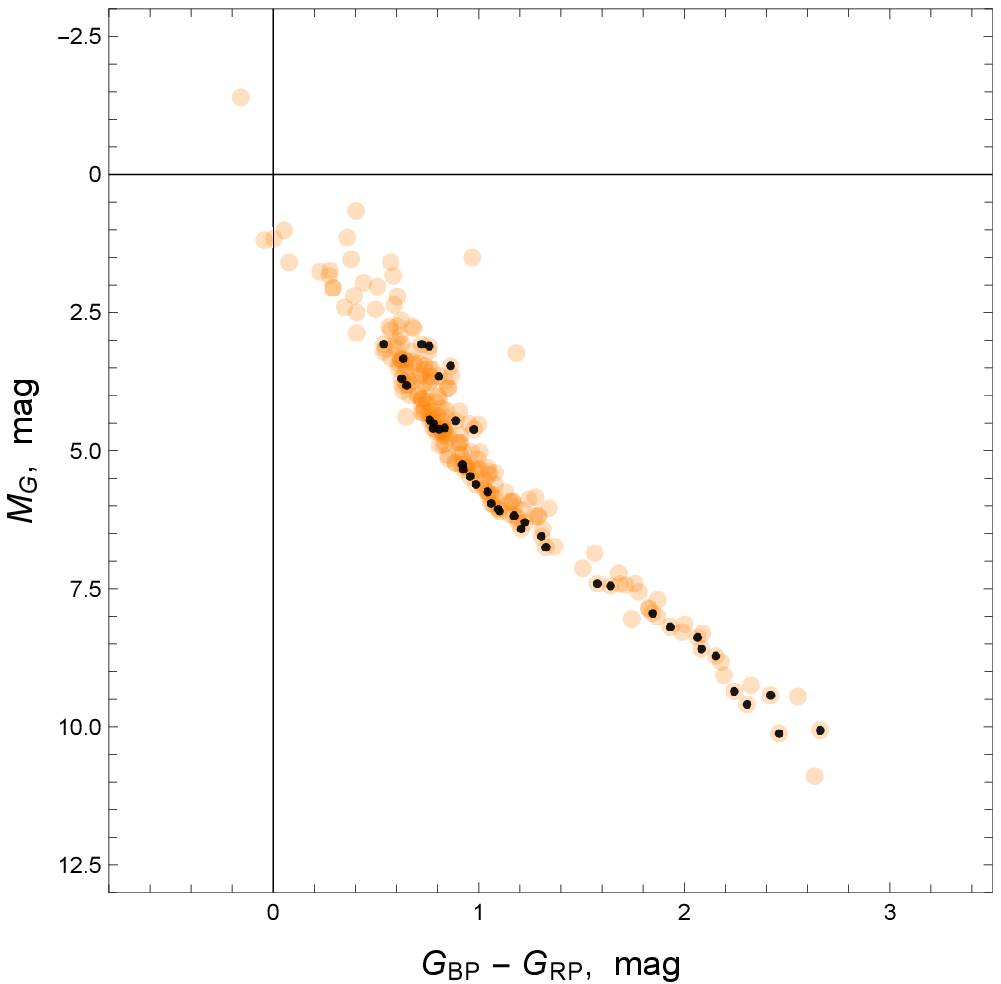}{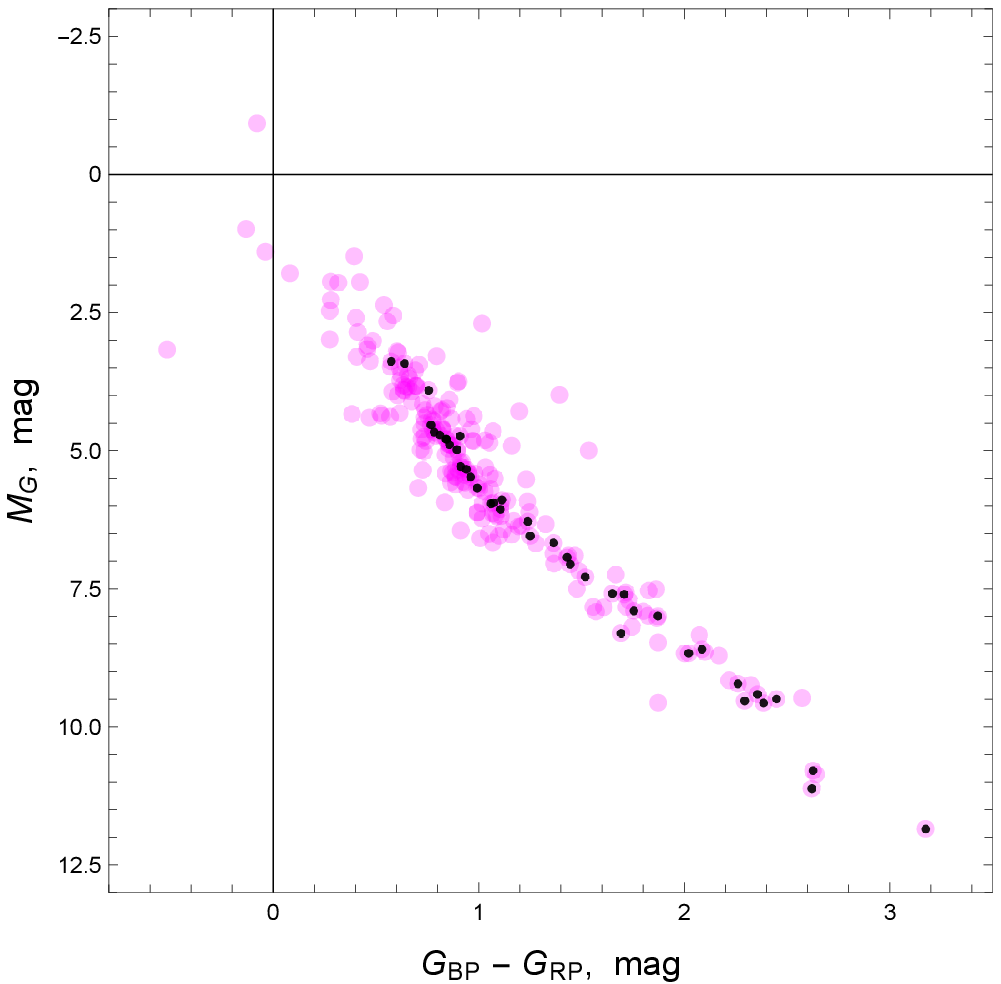}
\caption{Color-magnitude diagrams for primary (left) and secondary (right) components with astrometric mass ratios determined in this paper. The smaller black circles show the position of primary and secondary components
in pairs with wider separations, i.e., greater than $2.5\arcsec$.
\label{HR.fig} }
\end{figure}

Fig. \ref{HR.fig} shows the color-magnitude diagrams for the primary (left plot) and secondary (right
plot) components, computed with the $M_G$ absolute magnitudes and \gbp$-$\grp\ colors (whenever available)
from Gaia EDR3. Most of these components are main sequence dwarfs. A few stars may be slightly evolved,
possibly subgiants, and one secondary is remarkably blue and fairly luminous. This star is discussed among other
interesting cases in the following Section. However, photometric data from Gaia for resolved double stars
should be treated with caution. In particular, colors of the secondaries with small separation may be
strongly perturbed beyond the formal errors. The colors in Gaia are derived from a simple sum of the fluxes 
in the BP and RP spectra in windows of 2.1 arcsec across scan and 3.5 arcsec along scan. For separations below 1-2 arcsec, the photometry for the secondary is therefore compromised. The G magnitude is expected to be more
robust. As a results, at small separations, the two components get the same color, which can be seen in the
HR diagram of the secondaries if we color-code the data points according to the separation. These close secondaries are effectively dispersed mainly below the main sequence inheriting the blue colors of the primaries.
The black
circles in Fig. \ref{HR.fig} indicate the loci of components in pairs with separations greater than
$2.5\arcsec$, whose magnitudes are considered to be reliable. The main sequence is well defined for this smaller subset, with a mostly upward diffusion for some of the early dwarfs probably caused by post-MS evolution or hidden
tighter companions.

\section{Objects of note}
\label{obj.sec}
Some binaries stand out of the sample because of various kinds of oddities. We investigate a few of
them and perform literature search for those that caught our attention.
\subsection{HIP 89234 AB}
This well-known double star shows up in our catalog with the highest mass ratio $q=5.33_{-0.74}^{+0.96}$. Although
the wide error brackets and the quality flag 3 indicate a strongly perturbed solution, we would like to
understand what kind of perturbation can produce such a result. Theoretically, a stellar mass black hole
companion to the B component can generate a mass ratio of this magnitude. Such objects seem to be quite
rare and hard to find in the solar neighborhood even in much larger samples \citep{mato}. We have to carefully
inspect other, less exotic possibilities.

The system is listed as triple in WDS \citep{2021yCat....102026M}. The fainter companion resolved in Hipparcos
at a separation of $2.32\arcsec$, position angle (PA) 259.6 deg is a more distant tertiary. It is also
present in Gaia EDR3, albeit without \gbp, \grp\ magnitudes, with a separation of $2.103\arcsec$, PA$=270.7$
deg. This suggests the pair moves slowly counterclockwise on the sky with a long period. The inner pair 
(Aa and Ab) was
first resolved in 2008 at SOAR and named TOK 58. Only short segments of position measurements are available
for both pairs. Preliminary orbital fits suggest a period of 42 yr for the inner pair and $\sim 400$ yr for the
outer pair (A. Tokovinin 2021, priv. comm.). An orbital solution with a $P=324.8$ yr, $a=1.522\arcsec$, $e=0.628$
was proposed by \citet{ling} for the AB pair. Our bizarre solution for $q$ comes from the fact that
the observed vectors $v_1$ and $v_2$ in Eq. \ref{nor.eq} are $(-18.30,5.67)$ and $(-3.59,-0.54)$ mas
yr$^{-1}$, respectively, i.e., the observed orbital signal is much greater for the primary. It is
tempting to swap the cross-matched Gaia counterparts (which we had to do for two other systems as explained 
below), but the large magnitude difference precludes such an error in identification. The more likely
explanation is obtained from the character of the inner pair. This binary has moved almost linearly
between 2008 and now because of its inclination, which is close to $90\degr$. The Ab component has moved
mostly eastward on the sky at a proper motion of $\sim 60$ mas
yr$^{-1}$ relative to Aa. The observed reflex motion of Aa is roughly $-20$ mas yr$^{-1}$ in right ascension
due to the estimated mass ratio. Correcting for this inner pair motion takes out almost all of the
derived $v_1$. We conclude that in this case, our solution for $q$ suffers from the hierarchical trinary nature
of the system and physically perturbed Gaia proper motion measurements.

A similar case is found for HIP 62322 = $\beta$ Mus, an O-type supergiant and a member of the Sco-Cen
OB association. Our estimated $q=1.8$ of quality 3 likely originates from the orbital motion of the
tight inner pair that was discovered by \citet{2013MNRAS.436.1694R} separated by 45.6 mas with a magnitude
difference of 3.5. It may also be responsible for the large difference of 12.5 mas
yr$^{-1}$ between the proper motion in declination of the resolved A and B components. Being a very bright star,
$\beta$ Mus may also be astrometrically perturbed in Gaia because of the required special gating mode of measurements.

\subsection{HIP 2552 AB}
HIP 2552 = Gl 22 is an archetypical and well studied multiple system of late-type main sequence dwarfs,
for which our much perturbed solution  $q=4.3$ betrays the fast orbital motion of the inner, unresolved
components. A brief summary of the impressive observational history can be found in \citet{2008A&A...478..187D}.
The preliminary orbit of the outer pair (AB in our catalog) obtained by \citet{1973AJ.....78..935H}
from photographic measurements has been much improved over the decades and complemented with a tighter
inner orbit of the Aa and Ab components, as per WDS designation \citep{1999A&A...341..121S,2001A&A...368..572L}. 
The orbital periods of the inner Aa--Ab and the outer AB pairs are 15.6 and 223 yr, respectively. The inner
pair has an estimated semimajor axis of $0.51\arcsec$ and it passed its periastron in 2016. The estimated masses
are 0.377$M_{\sun}$ for Aa, 0.138$M_{\sun}$ for Ab, and 0.177$M_{\sun}$ for B. The presence of a small-amplitude
wobble in the motion of the B component suggests that it is a binary too \citep{2011Icar..215..712A} harboring
a companion with an estimated mass of $0.015M_{\sun}$, which can be a brown dwarf or a giant planet.
The astrometric picture is further complicated by the photometric variability and flare activity of the primary,
which is a UV Cet type active dwarf \citep{2014AcA....64..359T}. The A component is clearly disturbed in Gaia in both astrometry and $G$-band photometry. It has an \verb|ipd_frac_multi_peak| of 44\%, so the Ab component is 
in fact clearly visible to Gaia. Future Gaia releases will resolve this pair.
Since the primary is unresolved in either Hipparcos or EDR3, the
observed position is subject to a strong Variability Induced Motion (VIM) effect \citep[e.g.,][]{2016ApJS..224...19M} if it is observed during a flaring event. According to the WDS \citep{2021yCat....102026M},
the Ab component is fainter by 3.1 mag than the Aa component, and the latest observation puts them at
a separation of $0.4\arcsec$ in 2014. The photocenter of this pair observed by Gaia may be shifted by
$\sim 22$ mas from the true position of the primary. Flare amplitudes for M dwarfs of UV Cet type vary
in a wide range with the most energetic (but rare) events reaching several magnitudes. If an observed
flare with a peak amplitude of 0.1 mag takes place on the brighter primary component, the magnitude of
VIM is only about 2 mas. If the Ab component is responsible for the observed extra flux, the VIM excursion
is as high as 37 mas. This example shows that astrometric analysis of unresolved binaries with a variable
component may be complicated and ridden by ambiguity. Furthermore, the Hipparcos-epoch position is likely
perturbed too in the ``component" solution by the photocenter effects and by the considerable clock motion
of the inner pair. 

\subsection{HIP 4493}
This relatively less studied stellar system is listed in the WDS catalog with components A and B spanning
PA$=120\degr$, $\rho=0.4\arcsec$ in 1931 and PA$=53\degr$, $\rho=0.80\arcsec$ in 2008. A speckle interferometry
measurement in 2008.5461 places it at PA$=52.5\degr$, $\rho=0.822\arcsec$ \citep{2010AJ....139..743T}.
The pair has moved appreciably since then because Gaia EDR3 finds them at PA$=47.8\degr$, $\rho=0.836\arcsec$.
The secondary is fainter by 0.8 mag in $V_T$ \citep{2000A&A...356..141F}, however, the components have
nearly equal \gbp\ and \grp\ magnitudes in EDR3 while the $G$ magnitude of B is fainter by approximately
2 mag. This is the same effect of merged fluxes for close components we noticed in Section \ref{res.sec}.
The excess factor \verb|(Gflux/(BPflux+RPflux))| is 8.2, where something slightly higher than 1 is expected. 
Photometric data from \citet{2000A&A...356..141F} must clearly be preferred for this pair.
Another indicator that the Gaia solution is strongly perturbed comes from the RUWE parameter, which
is as high as 16.1 for A. It is not clear how much the Gaia data can be trusted for this object. Our
estimated $q=9.1$ comes from the fact that the long-term motion $\nu_2$ is close to the EDR3 proper motion of B
(within 1 mas yr$^{-1}$) while the corresponding values for A are quite different, so that $v_1=(6.3,4.9)$
mas yr$^{-1}$. We also note that the EDR3 proper motion of A differs from that of B by $(9.6,-3.4)$
mas yr$^{-1}$. Could this be a sign of fast orbital motion within the A component like in the previous
examples? A few circumstances cast doubt on this interpretation. The EDR3 parallax difference is 
significant at $2.4\sigma$ level, thus the pair just barely escaped our parallax consistency culling.
Furthermore, Gaia EDR3 lists a third, fainter and redder companion of $G=14.89$, \gbp$-$\grp$=2.14$ mag,
which is separated by $12.769\arcsec$ from A. With a parallax and a proper motion quite close to the
parameters of the B component, this star makes a legitimate member of a physical multiple system.
The data about this puzzling case suggest two routes of interpretation. The original A component may
be a rather improbable interloper leaving only a much wider binary pair BC. A more credible explanation is
that this is an outstanding borderline hierarchical triple system with an A component being strongly
perturbed in Gaia observations by an agent yet to be revealed.

\subsection{HIP 80017}
This is a known double star, which is undoubtedly a wide physical binary. The primary is an early and
blue dwarf of B9--9.5V type. \citet{2017MNRAS.471..770M} estimate a $T_{\rm eff}=8414$ $K$ and $L_{\rm bol}=
24.85\;L_{\sun}$ from astrometric and photometric data and spectral energy distribution models.
The pair appears to have barely moved between the Hipparcos and Gaia epochs changing the position angle
from $314.1\degr$ to $314.02\degr$. Given a separation of $1.224\arcsec$ and a parallax of 7.3 mas in Gaia,
a slow orbital motion is in line with our expectation, which also accounts for the low quality of our
mas ratio estimate $q=0.35^{+0.13}_{-0.09}$, SNR$=3.38$, $\eta=30\degr$. However, this unreliable
result draws much interest in view of the outstanding position of the secondary in the Hertzsprung-Russell
diagram in Fig. \ref{HR.fig} owing to its peculiar color \gbp$-$\grp$=-0.5\pm0.4$ mag. The star appears
to occupy the area of hot subdwarfs (sdO--sdB), which are dying stars in the post-AGB stage. These
stars are too blue to be regular dwarfs but still too luminous to be white dwarfs. To have an independent
mass estimate for a hot subdwarf would be interesting; however, this result should be taken with much caution.
The separation is small enough for the color-equalizing error (Section \ref{res.sec}) to take hold.
Indeed, the photometry excess factor is 5.4 and all its (rather few) transits are flagged as blended.
The formal error of 0.4 mag for
the \grp\ magnitude is unusual for stars of similar brightness. The more reliable Tycho-2 color from \citet{2000A&A...356..141F}
is much redder with $B_T=9.58\pm0.02$ and $V_T=9.02\pm 0.1$ mag, although it still seems too blue for
a K dwarf companion suggested by the magnitude difference and our estimated mass ratio.

\section{Discussion and future work}
\label{con.sec}

Using commonly accessible information from the ESA's Hipparcos and Gaia space missions, we were able
to process the entire cross-matched sample of resolved double stars and selected 248 high-probability binary
systems with statistically significant estimates of mass ratio. The purely astrometric method
proposed in \citet{mak21} is free of any astrophysical models and assumptions about the nature of
the observed stars. However, this method is founded on the assumption that the system under investigation is truly binary, i.e., there are no tighter subsystems perturbing the mean proper motions of the wide components. \citet{2010ApJ...720.1727L} estimate a rate of 45\% of tighter subsystems for a sample of
wide early M dwarf binaries. The rate of long-period binaries with components luminous in X-rays is 2.4 times 
higher than among fainter X-ray stars \citep{2002ApJ...576L..61M}, which indicates a significant rate of active short-period binaries
among widely separated systems. As a mitigating factor, the mean-epoch proper motions are not expected to be strongly perturbed by the inner pair's dynamics when the orbital period is much shorter than the mission
time span (33 months for Gaia EDR3).
 The output is presented as an online catalog with relevant information
copied from the Hipparcos DMSA and Gaia EDR3, as well as mass ratio estimates and their uncertainties
derived from a dedicated Monte Carlo simulation. The output is a small fraction of the initial 6306
doubles in the cross-matched sample with the required data. The attrition is mostly due to a high
rate of distant and wide physical pairs whose orbital motion is too slow to generate a significant
astrometric signal (i.e., a measurable change in components' proper motions) and the presence of
unrelated optical pairs. A closer look at a few peculiar outcomes reveals that the next significant
difficulty comes from hierarchical triple or higher multiplicity systems that are not resolved in either
DMSA or Gaia EDR3. Unless the orbital period of the inner subsystems is much shorter than the mission length,
the fast motion of the unresolved photocenter perturbs the mean-epoch proper motion of the corresponding
component. The effect on the estimated mass ratio depends on which component harbors a tight binary.
If it is the primary component, as in the triple system  HIP 89234, the additional apparent acceleration
perturbs mostly the short-term Gaia proper motion signal in the numerator of Eq. \ref{nor.eq}, leading to a
blown up estimate of $q$. If a binary is hidden in the secondary DMSA component, the extra astrometric
signal affects mostly the denominator resulting in a lower estimate for $q$. The former scenario is easier 
to catch, because due to the frequent occurrence of binaries with near-twin components, the resulting
$q$ may come up way above 1. Additional candidate triple systems can be identified upon a careful comparison
of the astrometric mass ratios with photometric estimates based on stellar evolution models.

The preponderance of wide, long-period systems in the DMSA sample is the greatest limitation to the
application of the astrometric method with the currently available data. The epoch difference of 24.75 yr
is too short to generate a measurable proper motion signal of binaries with period of the order of 1000
yr and longer. A next-generation space mission of absolute astrometry some 25--30 years after the Gaia mission
will give a boost to the signal-to-noise ratio on the known systems and will add thousands new targets
beyond the Hipparcos limiting magnitude. A Gaia follow-up in the near infrared \citep[Gaia-NIR,][]{hob,mca},
for example, is expected to widen the horizon of applicability including many objects of special\
astrophysical interest, such as degenerate stars, hot subdwarfs. Having two
epochs of absolute astrometry at the Gaia's level of accuracy separated by 25--30 years will lead to characterization
of millions of fainter and more distant binaries. In the shorter term perspective, the future Gaia data
releases (DR4 and DR5) will more than double the processed observational cadence, and a modified version of
this method for Gaia-only astrometric measurements will take it to a new level of accuracy of the empirical
binary star masses.

\section*{Acknowledgments}
The authors are grateful to the referee A. Tokovinin for valuable and constructive remarks.
This work made use of data from the European Space Agency (ESA) {\sc Hipparcos} astrometry satellite.
This work has made use of data from the European Space Agency (ESA) mission {\it Gaia} (\url{https://www.cosmos.esa.int/gaia}), processed by the {\it Gaia} Data Processing and Analysis Consortium (DPAC, \url{https://www.cosmos.esa.int/web/gaia/dpac/consortium}). Funding for the DPAC has been provided by national institutions, in particular the institutions participating in the {\it Gaia} Multilateral Agreement.
This work was (partially) supported by the Spanish Ministry of Science, Innovation and University (MICIU/FEDER, UE) through grant RTI2018-095076-B-C21, and the Institute of Cosmos Sciences University of Barcelona (ICCUB, Unidad de Excelencia 'Mar\'ia de Maeztu') through grant CEX2019-000918-M.


\begin{thebibliography}{99}

\bibitem[Andrade \& Docobo(2011)]{2011Icar..215..712A} Andrade, M. \& Docobo, J.~A.\ 2011, \icarus, 215, 712. doi:10.1016/j.icarus.2011.07.008
\bibitem[Arenou et al.(2018)]{are} Arenou, F., Luri, X., Babusiaux, C., et al.\ 2018, \aap, 616, A17. doi:10.1051/0004-6361/201833234
\bibitem[Docobo et al.(2008)]{2008A&A...478..187D} Docobo, J.~A., Tamazian, V.~S., Balega, Y.~Y., et al.\ 2008, \aap, 478, 187. doi:10.1051/0004-6361:20078594

\bibitem[ESA(1997)]{esa} ESA, 1997, The Hipparcos and Tycho Catalogues, ESA SP-1200

\bibitem[Fabricius et al.(2002)]{2002A&A...384..180F} Fabricius, C., H{\o}g, E., Makarov, V.~V., et al.\ 2002, \aap, 384, 180. doi:10.1051/0004-6361:20011822
\bibitem[Fabricius \& Makarov(2000a)]{2000A&A...356..141F} Fabricius, C. \& Makarov, V.~V.\ 2000a, \aap, 356, 141
\bibitem[Fabricius \& Makarov(2000b)]{2000A&AS..144...45F} Fabricius, C. \& Makarov, V.~V.\ 2000b, \aaps, 144, 45. doi:10.1051/aas:2000198
\bibitem[Fabricius et al.(2021)]{fab} Fabricius, C., Luri, X., Arenou, F., et al.\ 2021, \aap, 649, A5. doi:10.1051/0004-6361/202039834

\bibitem[Gaia Collaboration et al.(2016)]{pru} Gaia Collaboration, Prusti, T., de Bruijne, J.~H.~J., et al.\ 2016, \aap, 595, A1. doi:10.1051/0004-6361/201629272
\bibitem[Gaia Collaboration et al.(2021)]{bro} Gaia Collaboration, Brown, A.~G.~A., Vallenari, A., et al.\ 2021, \aap, 649, A1. doi:10.1051/0004-6361/202039657
\bibitem[Hershey(1973)]{1973AJ.....78..935H} Hershey, J.~L.\ 1973, \aj, 78, 935. doi:10.1086/111499

\bibitem[Hobbs et al.(2019)]{hob} Hobbs, D., Brown, A., H{\o}g, E., et al.\ 2019, arXiv:1907.12535
\bibitem[Lampens \& Strigachev(2001)]{2001A&A...368..572L} Lampens, P. \& Strigachev, A.\ 2001, \aap, 368, 572. doi:10.1051/0004-6361:20010024
\bibitem[Law et al.(2010)]{2010ApJ...720.1727L} Law, N.~M., Dhital, S., Kraus, A., et al.\ 2010, \apj, 720, 1727. doi:10.1088/0004-637X/720/2/1727
\bibitem[Ling(2016)]{ling} Ling, J.\ 2016, IAU Comm. 26, Circular 189
\bibitem[Makarov(2002)]{2002ApJ...576L..61M} Makarov, V.~V.\ 2002, \apjl, 576, L61. doi:10.1086/343088
\bibitem[Makarov et al.(2017)]{mafa} Makarov, V.~V., Fabricius, C., \& Frouard, J.\ 2017, \apjl, 840, L1. doi:10.3847/2041-8213/aa6af1

\bibitem[Makarov \& Tokovinin(2019)]{mato} Makarov, V.~V. \& Tokovinin, A.\ 2019, \aj, 157, 136. doi:10.3847/1538-3881/ab05e0

\bibitem[Makarov \& Goldin(2016)]{2016ApJS..224...19M} Makarov, V.~V. \& Goldin, A.\ 2016, \apjs, 224, 19. doi:10.3847/0067-0049/224/2/19
\bibitem[Makarov(2021)]{mak21} Makarov, V.~V.\ 2021, arXiv:2107.11274

\bibitem[Mason et al.(2021)]{2021yCat....102026M} Mason, B.~D., Wycoff, G.~L., Hartkopf, W.~I., et al.\ 2021, VizieR Online Data Catalog, B/wds
\bibitem[McArthur et al.(2019)]{mca} McArthur, B., Hobbs, D., H{\o}g, E., et al.\ 2019, \baas, 51, 118
\bibitem[McDonald et al.(2017)]{2017MNRAS.471..770M} McDonald, I., Zijlstra, A.~A., \& Watson, R.~A.\ 2017, \mnras, 471, 770. doi:10.1093/mnras/stx1433
\bibitem[Rizzuto et al.(2013)]{2013MNRAS.436.1694R} Rizzuto, A.~C., Ireland, M.~J., Robertson, J.~G., et al.\ 2013, \mnras, 436, 1694. doi:10.1093/mnras/stt1690

\bibitem[S{\"o}derhjelm(1999)]{1999A&A...341..121S} S{\"o}derhjelm, S.\ 1999, \aap, 341, 121

\bibitem[Tamazian \& Malkov(2014)]{2014AcA....64..359T} Tamazian, V.~S. \& Malkov, O.~Y.\ 2014, \actaa, 64, 359
\bibitem[Tokovinin et al.(2010)]{2010AJ....139..743T} Tokovinin, A., Mason, B.~D., \& Hartkopf, W.~I.\ 2010, \aj, 139, 743. doi:10.1088/0004-6256/139/2/743
\bibitem[Tokovinin et al.(2020)]{2020AJ....160..268T} Tokovinin, A., Petr-Gotzens, M.~G., \& Brice{\~n}o, C.\ 2020, \aj, 160, 268. doi:10.3847/1538-3881/abc2d6
\bibitem[Torres et al.(2002)]{2002AJ....124.1716T} Torres, G., Boden, A.~F., Latham, D.~W., et al.\ 2002, \aj, 124, 1716. doi:10.1086/342018

\end{thebibliography}
\end{document}